# Bridging the gap between diabetes and stroke in search of high clinical relevance therapeutic targets

Thierry Coppola, Sophie Béraud-Dufour, Patricia Lebrun and Nicolas Blondeau

**Informative title:**

Therapeutic targeting of common features of diabetes and stroke

**Affiliation:**

Université Côte d'Azur, CNRS, UMR7275 - IPMC, Sophia Antipolis, F-06560, France

Institut de Pharmacologie Moléculaire et Cellulaire

660 route des Lucioles, Sophia Antipolis

06560 Valbonne, FRANCE

**Correspondence** should be addressed to

Nicolas Blondeau (Blondeau@ipmc.cnrs.fr) and Thierry Coppola (Coppola@ipmc.cnrs.fr).

**ORCID of the authors:**

Dr Thierry Coppola: 0000-0003-1018-722X

Dr Sophie Beraud-Dufour: 0000-0002-0062-6842

Dr Patricia Lebrun: 0000-0001-9159-105

Dr Nicolas Blondeau: 0000-0001-5954-4094

**ACKNOWLEDGMENTS**

The authors thank all their past and present team members and collaborators who have contributed to the data discussed in the review. We would like to express our gratitude to Drs. Heurteaux and Mazella for their cordial support, valuable information, and guidance, which helped us in completing this review.

## ABSTRACT

Diabetes affects more than 425 million people worldwide, a scale approaching pandemic proportion. Diabetes represents a major risk factor for stroke, and therefore is actively addressed for stroke prevention. However, how diabetes affects stroke severity has not yet been extensively considered, which is surprising given the evident but understudied common mechanistic features of both pathologies. The increase in number of diabetic people, in the incidence of stroke in presence of this specific risk factor, and the exacerbation of ischemic brain damage in diabetic conditions (at least in animal models) warrant the need to integrate this comorbidity in pre-clinical studies of brain ischemia to develop novel therapeutic approaches. Therefore, a better understanding of the commonalties involved in the course of both diseases would offer the promise of discovering novel neuroprotective pathways that would be more appropriated to clinical situations. In this article, we will review the relevant mechanisms that have been identified as common traits of both pathologies and that could be to our knowledge, potential targets for both pathologies.

# 1- Introduction

Ischemic stroke is among the leading causes of mortality worldwide and of long-term cognitive disability and physical handicap. Several risk factors for ischemic stroke have been identified. They can be divided in two categories. Elderly, family history, gender, and ethnicity account for the non-modifiable risks, while high levels of blood cholesterol, triglycerides, sugar and pressure are acknowledged signs of modifiable risk factors that also include smoking and alcoholism. Given the coexistence of modifiable risk factors at up to 80% of stroke incidence (Stoger 2008), their control is one of the bases of current therapeutic approaches for stroke primary and secondary prevention. Biomarkers like hypercholesterolemia, hyperglycemia and hyperlipidemia are also indicators of obesity and/or diabetes. Following a continued increase in prevalence during the past decades, hypertension has been stabilized. In contrast, the prevalence of obesity and diabetes continues to expand in number in all industrialized countries. Considered as major but modifiable risk factors of stroke, obesity and diabetes have been considered almost exclusively within the domain of stroke prevention. However, how obesity and diabetes affect stroke severity has not yet been clearly clarified. While in some epidemiological studies obesity was associated with counter-intuitive improved stroke outcome (so-called "obesity paradox", (Scherbakov et al. 2011)), diabetes and hyperglycemia, its most notable feature, were associated to a higher mortality rate in stroke-patients (Kiers et al., 1992). Worth of note, the presence of Type 2 diabetes seems to vanish the paradoxical benefit of obesity on stroke survival (Adamopoulos et al. 2011). In addition, pre-clinical investigations on diabetic animals for evaluating efficacy of protective therapy against cerebral ischemic damage (as recommended by the Stroke Therapy Academic Industry Roundtable (STAIR) guidelines) (Fisher et al. 2009) acknowledged hyperglycemia as exacerbating factor of ischemic brain damage. The confounding effects of Type 2 diabetes were recently proposed as "negative conditioning" for stroke and acute myocardial infarction, as it alters clinical outcome and also interferes with treatment by tissue plasminogen activator in stroke patients and with ischemic conditioning in the setting of acute myocardial infarction (Jiang et al. 2019; Wider and Karin Przyklenk 2019). Overall preclinical and clinical evidences suggest that the impact of diabetes extends beyond increasing stroke frequency, but also exacerbates ischemic brain damage (Rehni et al. 2017; Rehni and Dave 2018).

Therefore the momentum is building in the field for scientists and physicians to gain insights in how diabetes exacerbates stroke complications and to evaluate the hypothesis that targeting diabetic aggravation of ischemic brain injury may offer more translational potential than conventional therapeutic approach.

**2- Stroke fact**

Apart from reperfusion therapies, for which most patients are still ineligible and only a limited number of those that do receive treatment experience a full recovery, stroke patients are currently left with an extremely limited repertoire of therapeutic options. New stroke incidence, deaths and disablement (50% of stroke survivors experience residual motor or cognitive/amnesic deficits severe enough to require assistance in daily living) are still growing. Historically, stroke research aimed to specifically target one deleterious mechanism of the ischemic cascade triggered by the occlusion of brain circulation. The specificity of neuronal death related to the pathophysiology of stroke rests in a massive and uncontrolled release of glutamate from neurons, the major physiological excitatory neurotransmitter in the mammalian brain. It leads to an overactivation of ionotropic glutamate receptors, predominantly the N-methyl-D-aspartate (NMDA) glutamate receptor subtype, that accounts for a lethal influx of calcium in the cell. However, employment of a mono-therapeutic approach targeting these specific features of ischemic stroke has utterly failed in Human clinical trials examining acute administration of a neuroprotective therapeutic after onset of stroke (Labiche and Grotta 2004; O'Collins et al. 2006). These failures led to diminished commitments in preclinical and clinical stroke research taking away the prompt perspective of finding a cure. Therefore, stroke is undoubtedly one of the most devastating diseases worldwide, exerting an enormous societal and economic burden (Moskowitz et al. 2010).

**3- Diabetes fact**

Type 2 diabetes accounts for 90% of diabetes cases nationwide and has been increasing at an alarming rate in association to the rise of obesity in the world. Diabetes affects more than 425 million people worldwide (Guariguata et al. 2014) and pharmaceutical companies are actively working to develop drugs for the broad market. Diabetes is a disorder of the assimilation, use and storage of sugars from food resulting in high blood glucose levels called hyperglycemia. Food is composed of lipids (fats), proteins (animal or vegetable proteins) and carbohydrates (sugars, starches). By passing through the intestine and then entering the bloodstream, nutrients provide energy to the body for its proper functioning. Eating leads to an increase in circulating level of sugar, and the corresponding carbohydrates will for the high majority be converted into glucose. As soon as the increase in blood sugar level is detected by pancreas, the pancreatic β cells, grouped into clusters called islets of Langerhans, start to secrete insulin. Insulin works like a switch, allowing glucose to penetrate the cells into the muscles and fat tissues for being processed and stored, and preventing glucose synthesis and release by the liver. Overall, it leads to a decrease of glucose amount in the blood. When energy or blood sugar levels drop, mainly between meals, another hormone called glucagon, releases glucose stored by the liver. To summarize, the balance of insulin and glucagon maintains stable blood sugar levels. While the alteration of this regulatory system of glucose homeostasis appears as a signature of diabetes, being sneaky and painless, the development of type 2 diabetes can go unnoticed for a long time. Indeed, it takes an average of 5 to 10 years from the appearance of the first hyperglycemia to diagnosis.

Repeated and prolonged hyperglycemia induces long-term damage to nerves and blood vessels throughout the body leading to blindness, foot disorders (that can lead to amputations), erectile dysfunction, kidney failure, heart attack and stroke. Therefore, treatments in type 2 diabetes aim to normalizing blood sugar levels. The first oral antidiabetic agents, sulfonylureas, which induce cell depolarization by inhibition of ATP-sensitive potassium ($K_{ATP}$) channels, were developed in the 1950s (Charbonnel and Cariou 2011). Their use still continues worldwide for triggering release of insulin from the pancreas. Belonging to the same class of insulin secretion potentiators, two other therapeutic molecules have been released on the market since 2008: dipeptidyl peptidase-4 (DPP-4) inhibitors and glucagon like peptide-1 (GLP-1) analogues used to promote insulin secretion without the risk of hypoglycemia observed with the sulfonylureas (Charbonnel and Cariou 2011). These two drug therapies take advantage of the stimulation of the "incretin effect" that is reduced or absent in type 2 diabetic patients, while in physiological conditions the insulinotropic actions of the incretin hormones account for two-thirds of the insulin normally secreted after food ingestion. Therefore drug-driven improved incretin-mediated augmentation of insulin secretion by the islets of Langerhans after a meal succeed in decreasing blood glucose levels in type 2 diabetic patients. Several long-lasting analogs of GLP-1 have been developed, for example exenatide and liraglutide. The other approach is to inhibit the enzyme that inactivates GLP-1.Thus several DPP-4 inhibitors that can be orally taken as tablets have been developed (Charbonnel and Cariou 2011). Nevertheless the number of people with diabetes is expected to pick above 650 million in the next 20 years.

## 4- Association between prevalence of Diabetes Mellitus, and Cardiovascular Disease and Stroke

Cardiovascular complications, such as coronary artery disease and stroke, account for the majority of the death of patients with type 2 diabetes. Indeed, nearly 80% of them die of a cardiovascular disease (CVD). In relation to a non-diabetic control population, the relative risk is multiplied by 2 to 6 (Beckman et al. 2002). In addition, patients with type 2 diabetes free from any coronary pathology have a similar risk to have a heart attack than a non-diabetic population who have previously had a coronary event (Haffner et al. 1998). All these elements indicate that the accelerated atherosclerosis found in type 2 diabetic subjects (that leads to considerable morbidity and mortality) needs an aggressive care strategy (Chiquette and Chilton 2002; Solomon 2003). With regard to stroke, it is known since a long time that diabetes increases the risk of stroke independently of its effect on blood pressure (Barrett-Connor and Khaw 1988; Burchfiel et al. 1994). In type 2 diabetic patients, the risk of suffering from stroke, as well as the risk of dying from stroke, is at least two times greater than in the normal population, independently of other known risk factors for CVD (Almdal et al. 2004). Among stroke survivors, diabetic patients also display a higher risk of long-term disability (Hankey et al. 2007). Interestingly, insulin resistance and impaired glucose tolerance, which are crucial steps in the course of diabetes development (known as pre-diabetic conditions), are also associated with the development of stroke incidence (Figure 1) (Kernan and Inzucchi 2011; Thacker et al. 2011). The interconnections between diabetes and stroke are so intense that both the Food and Drug Administration and the European Medicines Agency issued guidance on the necessity to evaluate the potential cross-actions of new drugs on diabetes and stroke (Castilla-Guerra et al. 2018). While anti-diabetics aimed at lowering circulating glucose have not been shown so far to reduce stroke incidence, it should be noted that recent randomized controlled trials have identified new anti-diabetics that may improve stroke outcomes (Chawla and Tandon 2017). There is therefore an emerging hope to discover new anti-diabetic drugs that will interfere with the course of stroke, reducing its incidence and/or its consequences. However, how diabetes increases the risk and/or the consequences of stroke is still weakly understood, and the interconnections between diabetes and stroke have not been fully investigated, which may be surprising given the common risk factors and mechanistic features between both pathologies.

### 5- Common risk factors

#### a. Behavioral risk factors: unbalanced nutrition and sedentarity

The role of the environment (Figure 1) in the development of diabetes, obesity and in the increase in their current prevalence has been extensively documented. Over the last few decades, in westernized populations, an increase of the caloric intake was monitored concurrently with a reduction of physical activity, explaining the development of obesity. Technological innovations and socio-demographic factors have also contributed to this fact. Factors such as lower prices and increased accessibility of nutritionally poor, but energy-dense foods have played their part, as well as the unrelenting marketing of this product. Fat intake has also increased significantly (saturated and trans fats) in recent years, promoting increased caloric intake (energy rich and flavourful content). Regarding carbohydrates, the last century witnessed an overall decrease in their consumption: previously accounting for more than 50% of the daily caloric intake, it nowadays represents less than 40%. Within carbohydrates, the consumption of complex forms has decreased, while the consumption of refined-sugars (mono and disaccharides) has considerably increased (Bleich et al. 2008).

In addition, a significant reduction in physical activity partly linked to a more sedentary urban lifestyle and an increase in time spent in front of television or other screens, also contributes significantly to the increase in obesity. This is particularly accurate for children and adolescents (Rennie et al. 2005). Such sedentary lifestyle increases all causes of death, doubles the CVD risk, diabetes and obesity, and increases the risk of high blood pressure, osteoporosis, lipid disorders, depression and anxiety. According to WHO, 60-85% of the world's population in both developed and developing countries has a sedentary lifestyle, making it one of the biggest public health problems of our time, although it is still not receiving enough attention (Charansonney and Despres 2010). It is also estimated that two thirds of children are not physically active enough, which will have serious consequences for their future health. Finally, overweight/obesity and physical inactivity have been closely linked to metabolic syndrome, an umbrella term used to describe a cluster of health conditions, including hypertension, hyperglycemia, hypertriglyceridemia, hypercholesterolemia and excessive waist fat, that occurrences are risk factors for stroke and diabetes.

#### b. Metabolic and cellular risk factors

Hypertension is very common in 30% of people with type 1 diabetes and 60% of people with type 2 diabetes. The concomitant manifestation of hypertension and diabetes synergistically increases the risk of stroke and cardiovascular accident (Figure 1). According to the ADA, the blood pressure target for diabetic patients is 130/80 mmHg (American Diabetes 2017; Chobanian 2017; Chobanian et al. 2003). Targeting this blood pressure level, that could be achieved with different antihypertensive agents, has largely contributed to the decreased of morbidity and mortality in people with diabetes since 1990 (Ford et al. 2007). Therefore, a large proportion of type 2 diabetes patients are on anti-hyperglycemic treatment plus antihypertensive drugs (often both aspirin and lipid-lowering agents that are mandated by the current standards of medical care). Although several antihypertensive treatment may be available, the interest in ACE inhibitors (angiotensin converting enzyme inhibitors) and ARBs (angiotensin II receptor blockers) is growing since they seem also to reduce the development of kidney injury and improve insulin action in diabetic patients (Kirpichnikov and Sowers 2002).

Hyperglycemia is closely related to increased morbidity and mortality in stroke and is recognized both as risk and aggravating factor in both diabetic and non-diabetic patients (Figure 1). Increment of glycated hemoglobin A1c blood level (HbA1c test evaluates the average level of blood sugar over the past months) is associated to an increased risk of first-ever ischemic stroke in both diabetic and non-diabetic patients (Mitsios et al. 2018). For several years, various studies have shown beneficial effects of treatments aimed at lowering blood sugar levels on ischemic risks and current guidelines strongly recommend reducing HbA1c levels to < 7% to reduce the risk of cardiovascular complications. Nevertheless in the ADVANCE study, intensive treatments aimed at reducing HbA1c levels (<6.5%; witnesses a suitable control of glycemia) did not necessarily show a decrease in cardiovascular risk (Group et al. 2008; Zoungas et al. 2014). Stricter therapeutic approaches with the objective of lowering HbA1c levels (<6%) (ACCORD study) have even indicated an increase in patient mortality. The same type of conclusion that glucose-lowering therapy did not significantly reduce cardiovascular events was first drawn in the Veterans Affairs Diabetes Trial (VADT) (Murata et al. 2009). Nevertheless ulterior evidence showed that this strategy had led to reduced CVD events in VADT participants with lower calcified coronary atherosclerosis (Reaven et al. 2009) and several limitations were noted in these randomized controlled trials (Mitsios et al. 2018). Overall, lessons from several studies indicate that intensive treatments for decreasing high blood sugar when initiated early lead to a reduction in cardiovascular risk.

Diabetes-related increase in intracellular glucose concentrations leads to the activation of deleterious metabolic pathways such as hexosamines and aldose reductase inducing the increased amount of reactive oxygen species, ROS and the depletion of antioxidant enzyme substrates. ROS are reactive molecules produced by aerobic metabolism. There is a link between the increase in ROS, under the control of NF-κB, and the production of inflammatory cytokines. In addition, there is an increase in production of glycated products and activation of protein kinase Cdelta (deltaPKC), which plays a role in apoptosis after cerebral ischemia (Perez-Pinzon et al. 2005). Altogether this suggests that diabetes manifestations including increased ROS and inflammation may silently weaken the brain before stroke and/or predispose the brain to exaggerated inflammatory and injury responses after stroke. Therefore it is not surprising that increased proinflammatory response due to diabetes is further exacerbated in response to stroke and probably is a major leader of the observed increased ischemic damage (Shukla et al. 2017).

Hypertriglyceridemia, hypercholesterolemia and excess in waist fat: With the increasing number of obese people worldwide (Poirier et al. 2006), we are witnessing a parallel progression in the other risks of chronic diseases. Obesity, mainly characterized by the accumulation of visceral fat, is often associated with an increase in low noise inflammation and elevation of serum factors such as cytokines and chemokines that can generate multiple deleterious effects, including CNS tissue damage induced by stroke (Allan and Rothwell 2003; Le Thuc et al. 2015). Excess ROS generated by hyperglycemia induces histone 3 methylation, which increases NF-κB expression (El-Osta et al. 2008). Obesity is often accompanied by dyslipidemia, increased very low density lipoprotein (VLDL) cholesterol, triglycerides, low-density lipoprotein (LDL) and decreased high-density lipoprotein (HDL) concentration. These changes are often associated with a predisposition to rapid and aggressive atherosclerosis (Watts and Playford 1998). LDL-lowering therapies -primarily statins- are commonly prescribed to reduce the risk of stroke. More interestingly, reduction in LDL-cholesterol and triglycerides and increase in HDL-cholesterol are known to be protective for stroke risk in patients with type 2 diabetes (Colhoun et al. 2004). A recent work showed that antihyperglycemic drugs like sodium-glucose co-transporter-2 (SGLT-2) inhibitor, dapagliflozin, might also suppress potent atherogenic LDL-cholesterol and increased HDL2-cholesterol, a favorable cardiometabolic marker (Hayashi et al. 2017). However, the approach for efficacious lipid modification in these high-risk individuals is somewhat complicated and deserves more attention.

Diabetes, obesity and the associated insulin resistance are related to low noise inflammation. This is characterized by the over-expression of cytokines produced by adipose tissue and activated macrophages (Hotamisligil 2006). Insulin resistance causes, partially by increasing Free Fatty Acid (FFA) in the blood, reduced glucose uptake in the liver, adipose tissues and muscles and increased hepatic neoglucogenesis. These sequential deregulation mechanisms lead to hyperinsulinemia as a compensation process of insulin resistance. The relationships between FFA, ROS, TNFα and other cytokines generate the expression of many genes associated with insulin resistance (Shoelson et al. 2006; Wellen and Hotamisligil 2005). The expression of a number of cytokines (IL1-β, CRP, Adiponectin) indicates a cellular response to stress. Thus, it is likely that this low noise inflammation may be a common causal factor in diabetes, insulin resistance, obesity and cardiovascular disorders. However, how the priming of these mechanisms affects stroke risk and severity has not yet fully considered despite evident mechanistic overlap and commonalty in signal transduction between both pathologies.

### 6- Mechanistic overlap / commonalty in signal transduction

The common features observed in different organs and cellular systems are often viewed as bases of the correlations existing between pathologies. This concept of conservation of cellular signaling and mechanisms between organs and cell types may also apply for diabetes and stroke. Many signaling pathways involved in the loss of function, in the regression of key organs or even in cell death are found both at the periphery and the central system. It is also interesting to note that the expression of the same genes is necessary for certain endocrine and neural functions. In addition, the cell dysfunctions are often driven by alteration of shared mechanisms such as the loss of expression of transcription factors such as CREB, NF-κB and kinases activated by second messengers, such as JNK, protein kinase A (PKA) and Ca2+/calmodulin kinase (CamK). Phosphorylation is a crucial post-translational modification of proteins, which is involved in a very large number of cellular processes (differentiation, division, proliferation, apoptosis, etc.) and particularly in signaling mechanisms.

#### a. Cyclic adenosine monophosphate (cAMP)

cAMP, the second most common and versatile messenger, controls a range of physiological processes such as regulated secretion (of neurotransmitters and peptide hormones), ion channel conductance, learning and memory, apoptosis and inflammation (Beavo and Brunton 2002). G protein-coupled receptors closely control the cellular content of cAMP through adenylyl cyclase and cyclic nucleotide phosphodiesterase (Hanoune and Defer 2001; Lugnier 2006). Protein kinase A (PKA) and the two EPAC proteins (cAMP-activated guanine nucleotide exchange factor for Ras-like GTPases) are the cAMP mediators. This signal transduction cascade can ultimately lead to the phosphorylation of CREB (Figure 2). The activation of the PKA phosphorylates the CREB Ser133 located in its inducible kinase domain. PKA pathway modulating CREB activity is implicated in survival and preservation of function of endocrine cells (Jhala et al. 2003) and neurons (Riccio et al. 1999). Preclinical studies indicate that new anti-diabetic therapies based on GLP1 receptor agonists may be capable, in addition to promote insulin secretion regulation, of facilitating the maintenance of endocrine function, boosting β cell proliferation while preventing their apoptosis through the activation of cAMP/PKA/CREB pathway (Lee and Jun 2014). Similarly in neurons, ligands capable of stimulating cAMP production increase the amount of P-CREB that is acknowledged as a multifaceted regulator of neuronal plasticity and protection (Figure 2) (Sakamoto et al. 2011).

### b. $Ca^{2+}$/calmodulin-dependent protein kinase (CaMK)

An increase in cytosolic calcium concentration generates signaling pathways that involve the multifunctional calcium-binding messenger protein calmodulin (CaM, Figure 2). The $Ca^{2+}$/CaM dependent kinase cascade (CaMKinase) consists of three kinases: CaM kinase kinase (CaMKK) and CaMK I and IV, which are phosphorylated by the activated CaMKK. These kinases are involved in survival and function maintenance processes (Soderling 1999). While expressed in all eukaryotic cells, these proteins are particularly abundant in the brain and immune cells. CaMKK and CaMK IV are located in both nucleus and cytosol, while CaMK I is only cytosolic. CaMK IV regulates transcription through the phosphorylation of transcription factors such as CREB. There is a crosstalk between CaMK and other signaling pathways like PKA and serine–threonine kinase Akt (also known as protein kinase B (PKB). The CaM Kinases cascade modulates apoptosis. For example, CaMKK is required for neuroprotection induced by the Akt pathway. Inversely, CaMKK can phosphorylate and activate the Akt pathway *in vitro* (Yano et al. 1998), that will inhibit apoptosis modulating the pro-apoptotic Bcl2 pathway (Soderling 1999). Interestingly, these protective effects have also been observed in the peripheral organs. The GLP1 receptor agonist, exendin 4, by activating CaMKK/CaMK IV increases glucokinase expression. This protein that prepares glucose for entering its metabolic pathways has a central role in the metabolic coupling between glucose concentration and insulin secretion (Murao et al. 2009). In addition, exendin 4, an efficient

GLP-1 analogue, controls the expression of the glucose transporter in β cells (Chen et al. 2011) *via* the CaMKK/CaMKIV pathway. In cytotoxic contexts, transcriptome analysis has shown that expression of genes encoding CaMKIIa and CaMKIV were suppressed under glucotoxic conditions (Sugiyama et al. 2011) and also under ischemic stroke in apoptotic neurons (Demyanenko and Uzdensky 2017).

### c. c-Jun N-terminal kinases (Jun Kinase)

While the members of c-Jun N-terminal kinases (JNKs, Figure 2) family represent promising therapeutic targets for protection against ischemic processes and diabetes, their number and variety of physiological implications render their therapeutic targeting extremely complex. Numerous studies have carefully addressed the role of these stress-activated kinases in both stroke and diabetes. JNKs are involved in DNA proliferation, apoptosis, motility, metabolism and repair. JNKs phosphorylate cJun Ser63 and 75 in response to stress signals (Behrens et al. 1999). JNKs are central in ischemia-induced apoptosis in brain (Borsello et al. 2003). Increased JNK activity has deleterious effects leading to ischemic cell death. Transient focal ischemia induces nuclear localization of p-JNK and increase in cJun phosphorylation shortly after reperfusion (Ferrer et al. 2003). JNK is also activated in cardiac ischemia and reperfusion (Fryer 2001). Inhibitory approaches have demonstrated the involvement of JNKs in ischemic-induced cell death (Waetzig and Herdegen 2005), and preclinical evaluation of specific targeting of the JNK signaling pathway for preventing increase in c-Jun activation has demonstrated protective effect against pancreatic β (Bonny et al. 2000) and neuronal (Borsello et al. 2003) cell death. In addition, experimental evidence suggests that pharmacological inhibition of JNK activity by new antidiabetics, including exendin-4, metformin and rosiglitazone − nonetheless acting through different pathways - may themselves be protective against brain ischemia in diabetic conditions (Shvedova et al. 2018). The JNK pathway, that increased activity was found in tissues from diabetics is linked to insulin resistance and type 2 diabetes (Wellen and Hotamisligil 2005). In the context of obesity and insulin resistance, JNK1 action seems predominant. JNK pathways are associated to transcription factor activity and abnormally elevated concentration of TNFα observed in obesity condition leads to JNK mediated phosphorylation at Ser 307 of IRS-1 (Hirosumi et al. 2002). While future search for selective JNKs inhibitors remains crucial, the concept that specific inhibitors for JNK would ameliorate insulin resistance and diabetes hold promises (Solinas and Becattini 2017).

### d. Transcription factor CREB

Among the transcription factors described for maintaining the neuronal phenotype, the cAMP response element binding protein (CREB, Figure 2) has been extensively studied. CREB is a mediator of response to neurotrophins such as the Nerve Growth Factor (NGF). Gene expression under the control of CREB is necessary for NGF-induced survival (Riccio et al. 1999). The activation of the cellular transcription factor CREB requires the activation of different upstream signaling pathways and kinases. But taken as a whole the increase in phosphorylated CREB is an acknowledged marker of cell survival. CREB plays a central role in survival processes, but also in many other physiological processes. Thus, it is not surprising that a decrease in its expression has considerable effects. CREB is not only activated by the stimuli necessary for growth and survival but also by cellular stresses, such as hypoxia and oxidative stress representing a form of cellular defense. Specifically, recovery from stroke is associated with increased plasticity around the infarct zone (Rivera-Urbina et al. 2015). This study indicates that there is a transcription under CREB control that is crucial for the modulation of neuronal excitability and in the structuring of cortical plasticity and memory. Recently, Caracciolo and colleagues (Caracciolo et al. 2018) have shown that the overexpression of CREB accelerates the recovery of motor deficits after stroke. In parallel, it has been shown that the CREB extinction in the insulin secreting endocrine β cell is associated with glucose intolerance in type 2 diabetes (Abderrahmani et al. 2006; Favre et al. 2011b). This decline in CREB activity is correlated with the emergence of a related gene transcript, ICER (inducible cAMP early repressor), that is the natural inducible CREB antagonist. The inappropriate expression of ICER explains the dysfunction of β cells and ultimately β cell death due to chronic hyperglycemia, hyperlipidemia and oxidized LDL (Salvi and Abderrahmani 2014). Chronic exposure of β cells to high glucose concentrations causes a decrease in CREB protein due to hyper ubiquitination and its subsequent degradation (Costes et al. 2009). Nevertheless, a fine regulation of the equilibrium in the expression level of CREB and ICER is required, as an uncontrolled CREB expression and activity will also lead to insulin-resistance. Indeed, in the insulin-resistant adipocytes of obese patients, reduction of ICER level elevates CREB activity, which activates the expression of the CREB-targeted gene activating transcription factor 3 (ATF3). ATF3 that is described as an adaptive response gene to stressful stimuli belongs to the ATF/cyclic AMP responsive element binding family of transcription factors. The ATF3 raise downregulates the expression of Glut4 in white adipose tissue (WAT), thereby contributing to systemic insulin resistance (Brajkovic et al. 2012; Favre et al. 2011a). Adding a level of complexity, it is worth of note that in obese condition the regulation of ICER differs among cell types, possibly leading to different modulation of CREB activity in cells other than adipocytes. With regard to stroke, at the central level, ATF3 is weakly expressed in normal condition but rapidly upregulated in response to ischemic injury. ATF3 expression in neurons is accepted as a neuroprotective step against stroke, while its role in heart failure is still controversial (Hunt et al. 2012; Zhang et al. 2011; Brooks et al. 2015).

### e. Transcription factor NF-κB

The family of NF-κB transcription factors includes a collection of dimeric proteins formed from the subunits p50, p52, RalA/p65, RelB and c Rel. In most resting cells, NF-κB is located in the cytoplasm, associated with its inhibitor IκB. In response to a wide range of stimuli, such as IL1β, TNFα, LPS or various forms of stress, IκB is phosphorylated by IKK kinase. After ubiquitination, IκB becomes a substrate for the proteasome, which releases a NF-κB dimer that can then enter the nucleus and regulate its target genes. This is part of the inflammatory response. The IKK/NF-κB signaling pathway is essential to connect inflammation with altered metabolism and decreased insulin action (Hotamisligil and Erbay 2008; Solinas and Karin 2010; Medzhitov 2008). Metabolic stress signals generate insulin resistance and/or a dysfunction of pancreatic β cells. In liver cells and adipocytes of obese people, NF-κB activation is induced in response to nutritional overload or cytotoxic factor production, allowing macrophage recruitment and subsequent inactivation of insulin receptor signaling.

Inflammatory reactions can be both beneficial and harmful to the brain, depending on the strength of their activation. The same can be said for NF-κB, which activation is required for getting the protective effect of several models of brain preconditioning (Blondeau et al. 2001), and is conversely crucial to mediate neuroinflammatory harmful effects following stroke. NF-κB is activated in neurons and glial cells under acute neurogenerative conditions such as stroke and trauma. NF-κB activation in neurons can promote their survival by inducing the expression of anti-apoptotic proteins such as Bcl2 and the enzyme Mn-superoxide dismutase (Chu et al. 1997; Tamatani et al. 1999). On the other hand, by inducing the production and release of inflammatory cytokines and reactive oxygen molecules, NF-κB activation in microglia and astrocytes can contribute to neuronal degeneration (Bruce et al. 1996; John et al. 2003). Therefore while the contribution of NF-κB pathways is complex, the idea of its therapeutic targeting in both diabetes and stroke is still the focus of intense investigations.

### 7- Concluding remarks

While consequences of diabetes are not anymore underestimated in stroke and as supported by the present review, it seems now intuitive that mechanistic overlaps are key issues in the genesis and consequences of stroke and diabetes. However, there is still a paucity of scientific and clinical data to direct the public and clinicians in this important area. Whereas, it is clear that more work should be devoted to understand how diabetes leads to an increased risk of stroke and an exacerbation of stroke damage, another area of interest has also emerged. This is the novel concept that therapeutic drugs aimed at targeting the overlapping pathways may benefit to both pathologies. We are now witnessing the beginning of a mindset switch from the FDA requesting cardiovascular outcome trials for testing putative side effects of glucose-lowering agents on cardiovascular risks, to the integration of cardiovascular outcome as readout for all new diabetes treatments to evaluate their capacity of significantly reducing cardiovascular risks in diabetic patients. This switch in mindset that these drugs may have benefits beyond lowering blood glucose could be of great value to discover new anti-diabetic treatments associated to the opportunity to reduce stroke risks and/or consequences. Such clinical fact echoes the Stroke Therapy Academic Industry Roundtable (STAIR) guidelines that have identified since long time the necessity of evaluating therapeutic candidates against stroke in animals presenting co-morbidities to closely reproduce stroke clinical scenario (Stroke Therapy Academic Industry 1999). As recently detailed in a nice review by Ashish K. Rehni, several animal models to study ischemic brain injury during diabetes may be available and their use should be spread to ensure generation of more clinically relevant data in the field (Rehni et al. 2017). Given the high incidence of stroke and exacerbation of ischemic brain damage in diabetic patients, the momentum is built to increase the development of animal studies to 1/ evaluate the importance of targeting the effect of diabetes on stroke outcomes, 2/ identify key pathophysiological mechanisms of diabetic enhancement of brain injury during stroke.

Research in this area would probably emerge of overlapping pathways and common drug targets against both pathologies.

**Funding**

This work was funded by the Centre National de la Recherche Scientifique - CNRS.

**Conflict of Interest**

The authors declare that they have no conflict of interest.

**FIGURE LEGEND**

**Figure 1: The hypnotic spiral linking diabetes, obesity, behavioral risk and ischemic stroke.**

A common view is that environmental risks (like nutrition and sedentarity) and metabolic pathologies (including obesity and diabetes) are interconnected and share common pathological features, including hyperglycemia and hypertension at the systemic level and increased inflammation and ROS at the molecular level. Overall each of these parameters individually or in association contribute to an hypnotic spiral toward increased risk of having a stroke and displaying exacerbated ischemic damage.

**Figure 2: Signaling pathways overlapping in diabetes and stroke and occurring in both β and neuronal cell.**

**A) Signaling pathways promoting cell survival.** Activation of specific receptors leads to the phosphorylation and activation of the transcription factor CREB (cAMP response element-binding protein) that in turn mediates transcription of genes involved in cell protection. Trophic factor activates PI3K (Phosphatidylinositol-4,5-bisphosphate 3-kinase) and ERK1/2 (extracellular-signal-regulated kinases) signaling pathways in cells, through the activation of tyrosine kinase receptor (receptors for trophic factors). Cyclic adenosine monophosphate signaling pathways are associated with GPCR (G protein coupled receptor activation) coupled with AC (adenylyl cyclase) protein. ATP (adenosine triphosphate); cAMP (cyclic adenosine monophosphate); Epac (exchange protein directly activated by cAMP); PKA (protein kinase A); MEK1/2 (MAP kinase/ ERK kinase 1); RSK (ribosomal protein S6 kinase); PDK1 (Phosphoinositide-dependent kinase 1); Akt (serine-threonine protein kinase); Rac (Rac family small GTPase 1).

**B) Signaling pathways promoting cell death/apoptosis:** the canonical NF-κB signaling pathway. TNFα (Tumor Necrosis Factor α) or IL-1β (interleukin-1 β) activates TNFR (Tumor Necrosis Factor Receptor) and IL-1R1(Interleukin-1 Receptor), respectively. Through a variety of adapter proteins and signaling kinases this leads to an activation of IKKβ in the IKK complex. This phosphorylation is a prerequisite for its subsequent polyubiquitination, which results in proteasomal degradation of IκBα. NF-κB then translocates to nucleus and activates target gene transcription leading to cell death. Il1-β binding onto its receptor (Il-1R1) leads to the activation of JNK (c-Jun N-terminal kinase) and also PKCδ (Protein Kinase C delta) pathways. MAPK (mitogen-activated protein kinase); MKK (MAPK kinase); MEKK1 (MAPK/Erk kinase kinase 1); IKK (IκB Kinase); NF-kB (Nuclear Factor-kB); TRADD (TNFRSF1A associated via death domain); TRAF2 (TNF receptor associated factor 2); IRAK (interleukin 1 receptor associated kinase 1).

Figure1	Click here to access/download;Figure;Fig1-Coppola.pdf

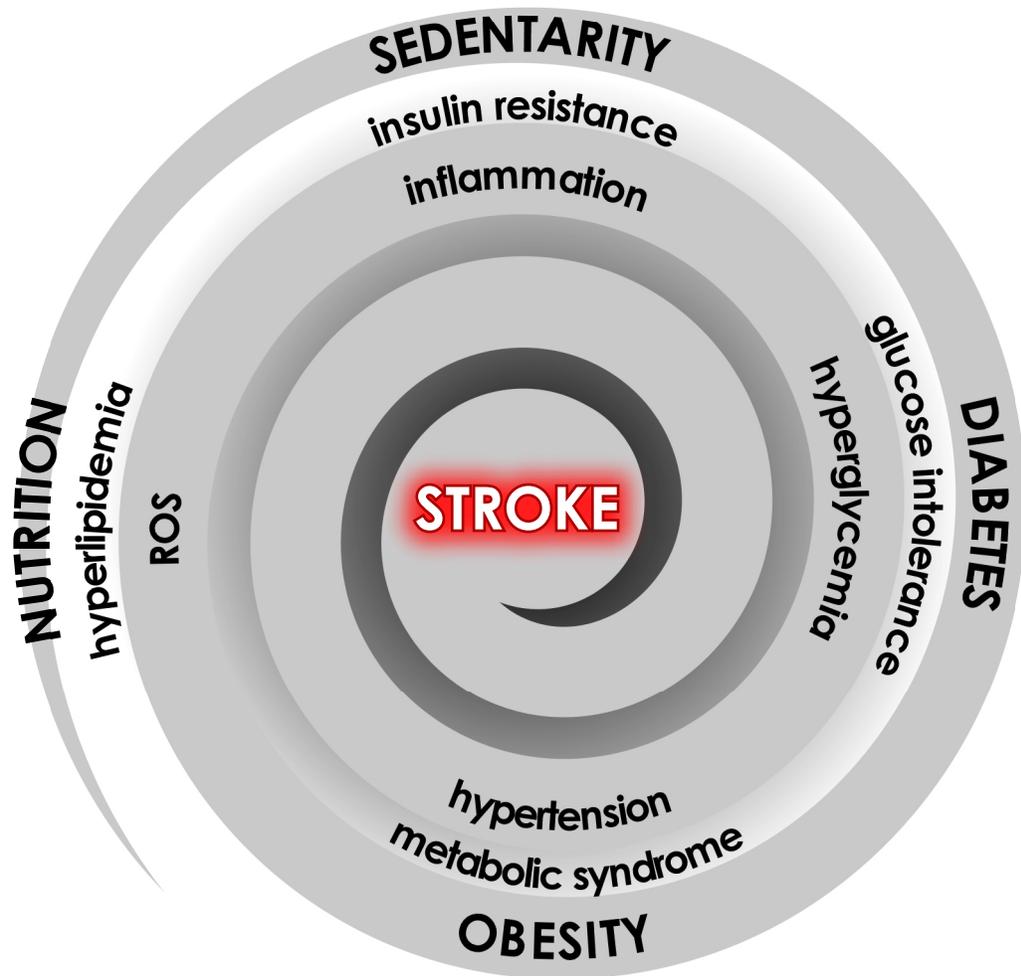

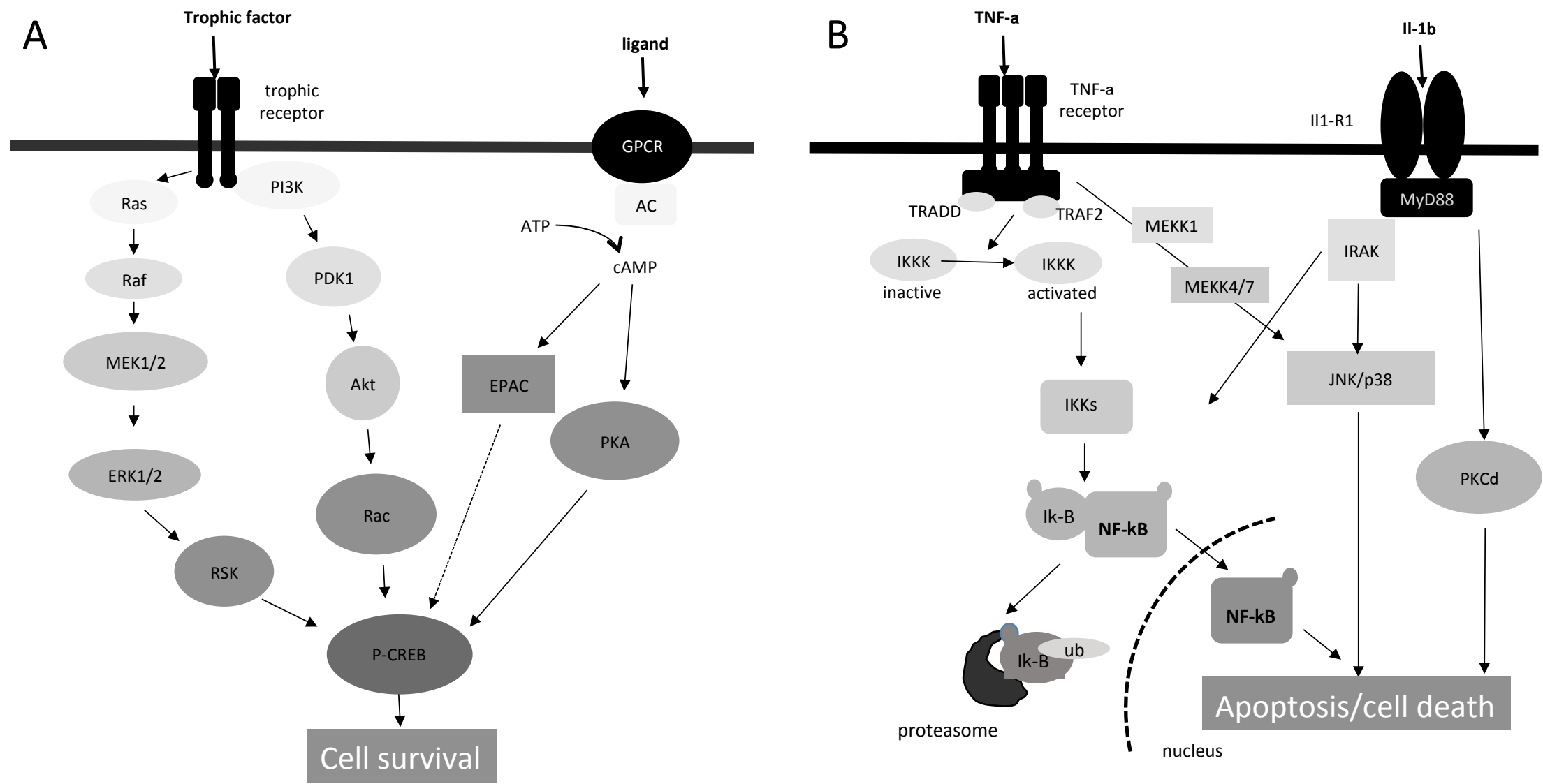